\documentclass[12pt,a4paper]{article}

\usepackage[utf8]{inputenc}
\usepackage[T1]{fontenc}
\usepackage{lmodern}
\usepackage[english]{babel}

\usepackage{geometry}
\usepackage{setspace}
\usepackage{microtype}
\usepackage{csquotes}
\usepackage{amsmath,amssymb}
\usepackage{enumitem}
\usepackage{xurl}
\usepackage{hyperref}

\hypersetup{
    colorlinks=true,
    linkcolor=blue,
    citecolor=blue,
    urlcolor=blue,
    pdftitle={The New Tractatus Program},
    pdfauthor={Mikołaj Sienicki and Krzysztof Sienicki},
    pdfsubject={Quantum Mechanics, Anti-Totalitarian Ontology, and Consciousness},
    pdfkeywords={Tractatus, Wittgenstein, quantum mechanics, consciousness, ontology, perspective}
}

\title{\textbf{The New Tractatus Program:}\\
Quantum Mechanics, Anti-Totalitarian Ontology, and Consciousness}

\author{%
\begin{tabular}{c}
Miko{\l}aj Sienicki\textsuperscript{1} and Krzysztof Sienicki\textsuperscript{2}\\[0.5em]
\begin{minipage}{0.92\textwidth}
\centering\small
\textsuperscript{1}Polish--Japanese Academy of Information Technology, ul. Koszykowa 86, 02--008 Warsaw, Poland, European Union.\\
\textsuperscript{2}Chair of Theoretical Physics of Naturally Intelligent Systems (NIS\textcopyright), Lipowa 2/Topolowa 19, 05--807 Podkowa Le{\'s}na, Poland, European Union.
\end{minipage}
\end{tabular}%
}

\date{\today}

\begin{document}

\maketitle

\begin{abstract}
This article reads two recent tractatus-style texts---Niccol{\`o}
Covoni and Carlo Rovelli's \emph{Tractatus Quanticus} and Miko{\l}aj
Sienicki and Krzysztof Sienicki's \emph{Tractatus de Conscientia}---together
with Jenann Ismael and Huw Price's interpretive introduction
\emph{Against Totalitarianism}, as parts of what may be called a New
Tractatus Program. The program is unified by a single pressure point:
the demand for a perspective-free description of reality, whether as a
final inventory of facts, a world viewed from nowhere, or a self placed
outside the physical order. On the relational interpretation of quantum
mechanics, \emph{Tractatus Quanticus} revises Wittgenstein's opening
claim by replacing the world as the totality of facts with the world as
what is the case from some perspective. \emph{Against Totalitarianism}
then makes the philosophical stakes of this move explicit by challenging
the idea of the world as one closed totality. \emph{Tractatus de
Conscientia} carries the same discipline into the theory of
consciousness, where conscious experience is treated neither as a
mysterious substance nor as mere behavior, but as an integrated,
temporally thick, operationally accessible perspective. Read together,
these texts point toward a post-classical philosophy of partiality:
reality is not abolished, but de-absolutized; consciousness is not
mystified, but constrained by access, coupling, memory, and evidence;
and the limits of language become inseparable from the limits of
perspective. The article does not claim that relational quantum
mechanics is the only viable interpretation of quantum theory. It
instead develops the philosophical consequences of the relational
reading. Its aim is programmatic rather than demonstrative: it does not
prove a new ontology, but organizes a family of arguments around
disciplined perspective-dependence.
\end{abstract}


\clearpage
\tableofcontents
\clearpage

% =========================================================
\section{Introduction: The Return of the \emph{Tractatus}}
% =========================================================

The \emph{Tractatus} has returned, but not as a monument. It has
returned as a way of thinking when the view from nowhere no longer
looks credible. In three recent texts---Covoni and Rovelli's
\emph{Tractatus Quanticus}, Ismael and Price's
\emph{Against Totalitarianism}, and Sienicki and Sienicki's
\emph{Tractatus de Conscientia}---Wittgenstein's old question about
the limits of meaningful speech is reopened under the pressure of
quantum mechanics and consciousness studies
\cite{Wittgenstein1922,CovoniRovelli2026,IsmaelPrice2026,SienickiSienicki2026}.
What emerges is not a repetition of early analytic philosophy, but a
new grammar of partiality. The world is no longer approached as a
completed inventory of facts visible from an absolute standpoint; facts
are facts from perspectives, perspectives are physically embodied, and
consciousness itself becomes intelligible only as a special kind of
integrated, temporally thick, operationally accessible perspective.

The phrase \enquote{New Tractatus Program} is used here to name this
shared movement. It does not designate a school, a doctrine, or a
finished system. It names a convergence: an attempt to inherit the
therapeutic and limiting function of Wittgenstein's early work while
replacing its classical background by post-classical physics and
operational theories of mind. The old \emph{Tractatus} asked what can
be said when the world is understood as the totality of facts. The new
tractatus-style texts ask what can be said when facts themselves are
perspectival, when no perspective-free description of reality is
available, and when consciousness must be approached through channels of
access, report, control, memory, and measurement.

For this reason, the article should be read as a programmatic synthesis
rather than as a finished formal theory. It does not try to derive a new
ontology from quantum mechanics, nor to solve consciousness by
definition. Its more modest aim is to clarify a common grammar of
perspective-dependence, coupling, record, and evidential limitation.

% ---------------------------------------------------------
\subsection{The \emph{Tractatus} as Form, Method, and Warning}
% ---------------------------------------------------------

The renewed appeal of the \emph{Tractatus} is not accidental. The
numbered-proposition form is not merely an aesthetic device. It
suggests compression, hierarchy, logical dependence, and a deliberate
refusal of ordinary argumentative looseness. A tractatus-style text
does not simply present conclusions; it attempts to discipline the
conditions under which conclusions may be meaningfully stated. In this
sense, the form itself already carries a philosophical claim: some
questions are confused not because they are too deep, but because they
are badly formed.

Wittgenstein's \emph{Tractatus Logico-Philosophicus} begins from the
famous claim that the world is everything that is the case, and that
the world is the totality of facts, not things
\cite{Wittgenstein1922}. Yet the work is also a warning against
mistaking the limits of a representational system for an object that
can be represented from outside. The subject, the world as a whole, the
limits of language, and the logical form of representation cannot be
treated as ordinary facts inside the world. They function instead as
limits. This is why the ladder of the \emph{Tractatus} must finally be
thrown away.

The new tractatus-style texts inherit this warning but redirect it. In
\emph{Tractatus Quanticus}, the warning is no longer only about the
limits of formal language. It becomes a warning against the
metaphysical temptation to imagine a complete, perspective-independent
description of the physical world. In \emph{Against Totalitarianism},
this temptation is named as the desire for a single totality: one
closed inventory of facts standing behind all perspectives. In
\emph{Tractatus de Conscientia}, the warning is applied to
consciousness: one must not demand a protocol-free identifier of
\enquote{what-it-is-like} when every scientific attribution of
consciousness requires some coupling between the investigated system and
an evidential channel.

Thus the tractatus form functions here in three ways. It is a
\emph{form}, because it organizes philosophical claims as compressed
propositions. It is a \emph{method}, because it separates meaningful
claims from pseudo-questions. And it is a \emph{warning}, because it
reminds us that the demand for an absolute standpoint may itself be the
source of philosophical illusion.

% ---------------------------------------------------------
\subsection{From the Limits of Language to the Limits of Perspective}
% ---------------------------------------------------------

The original \emph{Tractatus} is often read as a work about the limits
of language: what can be said clearly, what can only be shown, and what
must be passed over in silence. The New Tractatus Program preserves
this concern, but shifts its center of gravity. The relevant limit is
not only linguistic. It is perspectival, physical, and operational.

Covoni and Rovelli's decisive revision is to replace the classical
formula
\[
    \text{the world is the totality of facts}
\]
with the perspectival formula
\[
    \text{the world is everything that is the case from some perspective}.
\]

This should not be read as a subjectivist thesis. A perspective is not primarily a
mind, an ego, or a private inner theatre. In the relational quantum
mechanical setting, a perspective is a physical structure of interaction
and correlation. Facts are not free-floating atoms of an absolute
world-picture; they are values of variables in physical interactions.
To attribute a fact is to occupy, or to reconstruct, a perspective in
which that fact obtains.

This shift matters because it changes the status of total description. If facts are perspectival, then
the ideal of a complete description from nowhere becomes unstable. It
is not merely that human beings lack such a description. Rather, the
very idea of such a description may fail to correspond to anything
physically meaningful. The limit of language is therefore deepened into
a limit of perspective: what can be said must be said from somewhere,
by some system, through some coupling, within some space of possible
distinctions.

Ismael and Price make this point explicit. In their interpretation,
\emph{Tractatus Quanticus} does not merely add quantum mechanics to an
old logical picture. It transforms the picture. The classical world was
imagined as a shared public landscape whose objects possessed
properties independently of any observer. On the relational reading, quantum mechanics is interpreted as
removing this non-relational core from the description. There are
perspectives and relations among perspectives, but no further classical
inventory of intrinsic properties that all perspectives simply reveal
\cite{IsmaelPrice2026}. The view from nowhere is not hidden within the
relational picture; it is not part of that picture. This remains an
interpretive consequence of RQM, not a theorem forced on all viable
interpretations of quantum theory.

Reality does not disappear on this view. It means that reality is
de-absolutized. There are facts, but they are facts in perspectives.
There is objectivity, but it must be reconstructed through stable
relations among perspectives rather than presupposed as an
Archimedean inventory. There is knowledge, but it is embodied,
situated, and physically underwritten, in a sense continuous with situated-self
approaches to agency and perspective \cite{Ismael2006}.

A terminological caution is needed at the outset. The word
\enquote{perspective} will not be used in a single undifferentiated
sense. At minimum, the argument distinguishes a physical perspective
(\(P_{\mathrm{phys}}\)), constituted by interaction and correlation; an
epistemic perspective (\(P_{\mathrm{epist}}\)), constituted by
information available to a knower or system; a biological perspective
(\(P_{\mathrm{bio}}\)), constituted by organism-relative perception and
action; and a conscious perspective (\(P_{\mathrm{conc}}\)), constituted
by integrated, temporally extended, operationally accessible
experience. The central bridge claim is therefore not that all
perspectives are conscious, but rather that conscious perspectives form
a restricted and enriched subclass of physical perspectives.

For orientation, the following table summarizes how the four uses of
\enquote{perspective} function in the article:

\begin{center}
\scriptsize
\setlength{\tabcolsep}{3pt}
\begin{tabular}{p{2.2cm}p{2.5cm}p{2.8cm}p{3.0cm}p{2.4cm}}
\hline
\textbf{Domain} & \textbf{Perspective type} & \textbf{Fact or content type}
& \textbf{Evidence type} & \textbf{Main risk} \\
\hline
Quantum mechanics & \(P_{\mathrm{phys}}\) & value-in-interaction
& record, correlation, coupling & absolutism \\
Epistemology & \(P_{\mathrm{epist}}\) & accessible information
& inference, access, reconstruction & subjectivism \\
Biology & \(P_{\mathrm{bio}}\) & organism-relative salience
& perception, action, regulation & reductionism \\
Consciousness & \(P_{\mathrm{conc}}\) & appearance approached through access constraints
& report, control, memory, neural or causal trace & behaviorism or panpsychism \\
\hline
\end{tabular}
\end{center}

% ---------------------------------------------------------
\subsection{Three Recent Texts as One Philosophical Constellation}
% ---------------------------------------------------------

The materials considered here should therefore not be read simply
as independent literary exercises in Wittgensteinian style. Two of them
use the tractatus form directly, while Ismael and Price provide an
interpretive introduction that makes the ontological stakes explicit.
Together, they form a single philosophical constellation.

First, \emph{Tractatus Quanticus} provides the ontological and physical
basis of the program. It proposes that the world is not a totality of
absolute facts, but a structure of facts from perspectives. It also
physicalizes perspective: perspectives are not essentially mental, but
are embodied in the ways systems interact and in the information
available through those interactions \cite{CovoniRovelli2026}.

Second, \emph{Against Totalitarianism} provides the philosophical
interpretation of this move. It clarifies that the target is not
realism as such, but the totalitarian temptation in ontology: the
belief that there must be one final, complete, perspective-independent
world-description. On this reading, quantum mechanics does not merely complicate the old
picture. It is interpreted, within RQM, as removing the metaphysical
ground on which that picture rested \cite{IsmaelPrice2026}.

Third, \emph{Tractatus de Conscientia} extends the same grammar to
consciousness. Its central problem is not whether consciousness is a
mysterious non-physical substance, nor whether it can be reduced to
outward behavior. Its problem is to specify the operational and
structural conditions under which a physical system can be said to
possess a conscious point of view. It therefore distinguishes
appearance, access, and formal structure; treats conscious episodes as
temporally extended and integrated; and insists that claims about
consciousness require explicit evidential channels
\cite{SienickiSienicki2026}.

The three texts therefore stand in a natural sequence:
\[
\begin{array}{rcl}
\emph{Tractatus Quanticus} &:&
\text{facts are perspectival;} \\[2mm]
\emph{Against Totalitarianism} &:&
\text{there is no final world-totality behind perspectives;} \\[2mm]
\emph{Tractatus de Conscientia} &:&
\text{consciousness is a special, operationally constrained perspective.}
\end{array}
\]

The sequence also brings out a distinction on which the whole
program depends:
\[
    \text{all conscious perspectives are physical perspectives,}
\]
but
\[
    \text{not all physical perspectives are conscious perspectives.}
\]

A detector, a measuring device, or a particle may instantiate a
physical perspective in the broad relational sense. A conscious
perspective, however, requires additional organization: integrated
distinctions, temporal thickness, access, memory, control, and possible
report. The New Tractatus Program therefore generalizes perspective at
the level of physics, but restricts and enriches it at the level of
consciousness.

% ---------------------------------------------------------
\subsection{Main Thesis of the Article}
% ---------------------------------------------------------

The main thesis of this article is that the three texts, taken together,
constitute a post-classical and post-Wittgensteinian program of
philosophical clarification. Their common target is the fantasy of an
absolute standpoint: a world seen from nowhere, a total inventory of
facts, a self outside the world, or a private essence of consciousness
independent of every possible protocol of access.

Against this fantasy, the New Tractatus Program advances four linked
claims.

First, facts are perspectival. A fact is not thereby unreal; rather,
its reality is articulated within a perspective constituted by physical
relations.

Second, perspectives are physical. They are not necessarily mental,
biological, or personal. They are embodied in interactions,
correlations, variables, and informational structure.

Third, totality is not an object of knowledge. The demand for a single
final description of reality repeats the metaphysical error that the
original \emph{Tractatus} already exposed: the attempt to step outside
the world in order to describe the world as a bounded whole.

Fourth, consciousness is not an exception to this grammar. It should be
understood neither as a supernatural addition to physics nor as mere
behavior. It is better treated as a special kind of perspective: an
integrated, temporally thick, operationally accessible organization of
distinctions, stabilized enough to guide action, memory, control, and
sometimes report.

The slogan of the article may therefore be stated as follows:
\begin{quote}
The New Tractatus Program does not deny reality; it denies that reality
must be available as a single, final, perspective-free totality.
\end{quote}

In this sense, the return of the \emph{Tractatus} is not a return to
logical atomism. It is a return to philosophical discipline after the
collapse of the view from nowhere. What returns is the ladder, but the
ladder now leans against a different wall: quantum mechanics,
anti-totalitarian ontology, and the operational study of consciousness.

% =========================================================
\section{The Original Problem: World, Fact, Self, and Silence}
% =========================================================

The New Tractatus Program can only be understood against the original
problem posed by Wittgenstein: how can language describe the world
without pretending to stand outside it? The problem is at once logical,
metaphysical, and therapeutic. It concerns the relation between world,
fact, subject, representation, and silence. The point of this section
is therefore not to provide a full interpretation of Wittgenstein's
\emph{Tractatus}, but to isolate those features of the original problem
that later return, transformed, in quantum perspectivalism and in an
operational theory of consciousness.

% ---------------------------------------------------------
\subsection{Wittgenstein's World as the Totality of Facts}
% ---------------------------------------------------------

Wittgenstein's \emph{Tractatus Logico-Philosophicus} opens with one of
the most compressed ontological gestures in twentieth-century
philosophy:

\[
    \text{The world is everything that is the case.}
\]

This is immediately sharpened into the claim that the world is the
totality of facts, not of things \cite{Wittgenstein1922}. The
displacement is decisive. Reality is not first presented as a warehouse
of objects, but as a structured field of obtaining states of affairs.
Objects matter because they enter into possible configurations; facts
matter because they determine what is the case.

This move belongs to the broader logical-analytic background of Frege
and Russell. Frege had already shifted philosophy away from the
psychology of ideas toward the objective structure of sense and
reference \cite{Frege1892}. Russell, especially in his logical atomist
period, treated the world as analysable into facts that propositions
could represent \cite{Russell1918}. Wittgenstein inherits this
logical-atomist background, but radicalizes it. He does not merely ask
what facts there are; he asks what must be the case for language to
picture facts at all.

This is why the original \emph{Tractatus} matters for the new program.
If the early Wittgensteinian formula is

\[
    \text{world} = \text{totality of facts},
\]

then the new tractatus-style revision asks whether facts can still be
understood as absolute. \emph{Tractatus Quanticus} answers negatively:
facts are not abolished, but relocated within perspectives. The
transition is therefore not from facts to fiction, but from absolute
facts to perspectival facts.

% ---------------------------------------------------------
\subsection{The Impossibility of the View from Nowhere}
% ---------------------------------------------------------

The second essential problem is the impossibility of describing the
world from outside. If the world is everything that is the case, then
there is no external position from which the world can be surveyed as
one more object. Russell saw this difficulty clearly in his introduction
to the \emph{Tractatus}: whatever can be said must concern bounded
regions of the world, not the world as a completed object placed before
an external observer \cite{Russell1922}.

This is the first form of anti-totality. The totality of the world is
not one more fact among facts. To make it into such a fact would require
a position outside the totality, but that position is precisely what the
concept of totality excludes. The world as whole is not hidden from us
in the way a distant planet may be hidden. It is unavailable as an
object because any object of description would already belong to the
world.

Thomas Nagel later gave a famous name to the contrary temptation: the
\enquote{view from nowhere} \cite{Nagel1986}. In Nagel, the expression
names the aspiration toward an increasingly objective conception of
reality. In Wittgenstein's early work, however, the aspiration becomes
self-undermining when it tries to describe the totality from no
perspective at all. The New Tractatus Program inherits precisely this
diagnosis, but relocates it. The issue is no longer only whether
language can represent the world as a whole; it is whether physics
itself permits a perspective-free inventory of facts.

A useful way to state the continuity is therefore this:

\[
    \text{Wittgenstein: no language from outside the world;}
\]

\[
    \text{New Tractatus Program: no physics from outside perspective.}
\]

This does not eliminate objectivity. It rejects only the fantasy that
objectivity must take the form of an impossible, non-situated gaze.

% ---------------------------------------------------------
\subsection{The Self as Limit, Not Object}
% ---------------------------------------------------------

The third problem is the status of the self. The metaphysical subject,
for Wittgenstein, is not an item inside the world. It is not found among
facts. It is better understood as the limit of the world, the point from
which a world is encountered but which cannot itself be represented as
one object among others \cite{Wittgenstein1922}.

This idea has a complex ancestry. Kant had already distinguished the
empirical self, which can be studied as an object of inner sense, from
the transcendental unity of apperception, which is not itself another
empirical object \cite{Kant1781}. Schopenhauer, who influenced the
young Wittgenstein, similarly treated the subject as the condition of
the world as representation rather than as an object inside that world
\cite{Schopenhauer1818}. Mach, another important background figure,
criticized the substantial ego and described the self as a relatively
stable complex rather than as a metaphysical entity \cite{Mach1886}.

Wittgenstein's contribution is to translate this problem into the
grammar of representation. The \enquote{I} imagined as a pure observer
cannot be one of the things observed. Elizabeth Anscombe later attacked
the idea that the first-person pronoun functions as an ordinary
referring expression, thereby giving a related grammatical diagnosis of
the philosophical illusion surrounding the self \cite{Anscombe1975}.
The self is not simply another name for a hidden object.

This point is essential for consciousness theory. A theory of
consciousness must not reintroduce the self as a small metaphysical
spectator behind experience. If the self is to enter scientific theory,
it must enter as structure, function, or role: memory, access, control,
self-indexing, and integration. This is exactly why
\emph{Tractatus de Conscientia} treats the self not as a hidden
substance, but as a dynamical index binding episodes across time.

% ---------------------------------------------------------
\subsection{The Ladder, Nonsense, and Philosophical Therapy}
% ---------------------------------------------------------

The final problem is silence. The \emph{Tractatus} ends with the famous
command that whereof one cannot speak, thereof one must be silent
\cite{Wittgenstein1922}. But this silence is not simple ignorance. It is
not the silence of someone lacking information about an ordinary fact.
It is the silence required when a question exceeds the grammar that
would make an answer meaningful.

This is why the ladder image is so important. The propositions of the
\emph{Tractatus} are to be used and then recognized as nonsensical in a
special sense: they help the reader see the limits of sense, but they
cannot themselves state those limits from a position beyond them. Later
readers have disagreed sharply about how to understand this. Hacker
emphasizes the therapeutic and grammatical role of Wittgenstein's
method \cite{Hacker1986}; Diamond and Conant push a more radical
\enquote{resolute} reading, according to which the book's apparent
metaphysical theses are themselves part of what must be overcome
\cite{Diamond1991,Conant2002}. Carnap, from a different angle, treated
many metaphysical statements as pseudo-statements produced by misuse of
language \cite{Carnap1932}.

For the present article, the important point is not to settle these
interpretive disputes. It is to observe that the New Tractatus Program
inherits the therapeutic structure. Its task is not to add new
metaphysical objects called \enquote{perspectives}, \enquote{worlds},
or \enquote{consciousnesses}. Its task is to discipline the conditions
under which such terms can be used without illusion.

Thus silence returns in a new form. In quantum foundations, we must be
silent about facts for which no perspective supplies determinate
conditions of attribution. In consciousness studies, we must be silent about a private essence of
experience where no specified evidential coupling, access channel, or
operational distinction can identify it.
This is not a defeat of thought. It is a discipline of thought.

The original problem can therefore be summarized as follows:

\[
\begin{array}{rcl}
\text{world} &:& \text{not a thing outside facts;} \\[1mm]
\text{fact} &:& \text{not an isolated object, but what is the case;} \\[1mm]
\text{self} &:& \text{not an object inside the world, but a limit or role;} \\[1mm]
\text{silence} &:& \text{not ignorance, but the boundary of meaningful assertion.}
\end{array}
\]

The New Tractatus Program begins exactly here. It asks what becomes of
world, fact, self, and silence once the classical background of the
original \emph{Tractatus} is replaced by quantum mechanics,
anti-totalitarian ontology, and an operational theory of consciousness.

% =========================================================
\section{\emph{Tractatus Quanticus}: Quantum Mechanics and Perspectival Facts}
% =========================================================

\emph{Tractatus Quanticus} is the first decisive step in the New
Tractatus Program. It does not merely imitate Wittgenstein's form. It
revises the ontological grammar of the original \emph{Tractatus} under
the pressure of quantum mechanics. Where Wittgenstein begins with the
world as the totality of facts, Covoni and Rovelli begin with facts
from perspectives. The central shift is therefore:

\[
    \text{facts as absolute constituents}
    \quad \longrightarrow \quad
    \text{facts as perspectival physical events}.
\]

This is not a rejection of facts. It is a rejection, internal to the
relational interpretation of quantum mechanics, of the assumption that
facts must belong to a perspective-independent inventory of reality.
The claim should therefore be read interpretively rather than as a
settled theorem of quantum physics. Other interpretations of quantum
mechanics may preserve different forms of realism; the present article
develops the philosophical consequences of the relational reading.

% ---------------------------------------------------------
\subsection{\enquote{The World is Everything that is the Case from Some Perspective}}
% ---------------------------------------------------------

The opening thesis of \emph{Tractatus Quanticus} deliberately echoes
and revises Wittgenstein:

\[
    \text{The world is everything that is the case from some perspective.}
\]

This formula keeps the Wittgensteinian priority of facts over things,
but removes the classical assumption that facts are simply there,
absolutely, waiting to be collected into a totality. A fact is always a
fact \emph{from} or \emph{in} a perspective. The question whether there
are facts outside every perspective is therefore not an empirical
question with a hidden answer. It is a question whose grammar has
failed.

The important point is that this is not anti-realism. The world does
not disappear. Rather, the world is articulated through physically
situated facts. Reality remains, but the fantasy of a final,
perspective-free inventory of reality is abandoned.

% ---------------------------------------------------------
\subsection{Perspectives are not Minds: The Physicalization of Perspective}
% ---------------------------------------------------------

The word \enquote{perspective} might suggest subjectivity, experience,
or mental life. \emph{Tractatus Quanticus} explicitly blocks this
reading. A perspective is not essentially a mind, a person, a conscious
observer, or a biological organism. It is a physical relation among
systems. In this sense, a perspective is closer to a reference frame in
relativity than to a private psychological viewpoint
\cite{Rovelli1996,CovoniRovelli2026}.

This is the crucial move: perspective is physicalized. A detector, a
thermometer, a particle, a laboratory, or a brain may all instantiate
perspectives in the broad relational sense. What matters is not
mentality but interaction, correlation, and the availability of facts
relative to a system.

Thus the New Tractatus Program begins by generalizing perspective
beyond consciousness:

\[
    \text{perspective} \neq \text{mind}.
\]

This distinction later becomes essential for consciousness theory,
because it allows us to say that all conscious perspectives are
physical perspectives, without saying that all physical perspectives
are conscious.

% ---------------------------------------------------------
\subsection{Facts as Values of Variables in Interaction}
% ---------------------------------------------------------

The next move is to translate the language of facts into the language
of physical variables. In \emph{Tractatus Quanticus}, a property is a
fact expressed as the value of a variable. If a variable \(A\) has value
\(a\), the corresponding fact may be written:

\[
    A = a.
\]

Here \(A=a\) should not be read as a hidden intrinsic value possessed
prior to interaction, but as shorthand for an event, outcome, or record
relative to a specified physical interaction.

But the value of \(A\) is not an intrinsic possession of a system
independently of interaction. Variables express ways in which systems
manifest themselves to other systems. A value is therefore not a
metaphysical label attached to an object, but the outcome of a possible
or actual physical coupling.

This is where the quantum revision becomes sharp. In classical physics,
one may imagine that all relevant variables possess definite values
whether or not they are measured. In quantum mechanics, this assumption fails for non-commuting
observables and context-dependent value attributions. Not all variables
can consistently be assigned simultaneous determinate values in a given
perspective. Some questions, when asked without specifying a
measurement context or perspective, are not merely unanswered. They may
be badly formed.

The analogy is useful but limited:

\[
    \text{velocity without reference frame}
    \quad \sim \quad
    \text{quantum value without perspective}.
\]

To ask for the value of a quantum variable without specifying the
perspective in which the value is to obtain is to repeat a classical
metaphysical mistake. The analogy should not be pressed too far,
however. Relativistic frame-dependence and quantum perspectivality are
not the same physical phenomenon. Quantum value-attribution involves
non-commuting observables, measurement context, probabilistic state
assignment, and record-producing interactions, not merely a change of
coordinates.

% ---------------------------------------------------------
\subsection{Information, Probability, and Mutual Information}
% ---------------------------------------------------------

The perspectival account of facts requires an account of information.
If a fact is a fact in a perspective, then knowledge is not a detached
mirror of the world. It depends on physical relations between systems.
\emph{Tractatus Quanticus} therefore treats information as correlation,
not as an immaterial ingredient added to physics. This does not mean,
however, that knowledge is identical with mere correlation. Epistemic
knowledge requires interpretation, reliability, and conditions of
access; physical information is its material precondition, not its full
philosophical equivalent.

As a schematic classical model, let \(A=\{a\}\) and \(B=\{b\}\) be two
sets of possible facts, and let \(p(a,b)\) be their joint probability.
If \(p(b)>0\), then the probability of \(a\) conditional on \(b\) is

\[
    p_b(a) = p(a|b) = \frac{p(a,b)}{p(b)}.
\]

A record \(b\) may support the attribution of \(a\) when this
conditional probability is close to one, but a high-probability
inference is not automatically the same thing as a recorded fact. The
threshold, the measurement context, and the relevant operational
conventions must be specified.

Similarly, mutual information may be written schematically as

\[
    I(A:B),
\]

but this notation is deliberately heuristic here. In a fully quantum
treatment one must specify the states, observables, measurement
operators, records, and the relevant information measure. Classical
Shannon mutual information \cite{Shannon1948}, accessible information, and quantum mutual
information are not interchangeable. The quantum mutual information,
for example, is usually written as

\[
    I(A:B)=S(A)+S(B)-S(AB),
\]

where \(S\) is the von Neumann entropy. The formulas in this subsection are therefore not intended to derive
the ontology. They are schematic markers of the kind of formal machinery
that a mature version of the program would require. The philosophical
point is not that one formula solves the problem, but that facts,
records, and knowledge claims must be tied to physically instantiated
correlations and to the protocols by which they become available.

% ---------------------------------------------------------
\subsection{Measurement as Ordinary Physical Coupling}
% ---------------------------------------------------------

On this view, measurement is not a mysterious exception to physical
evolution. It is an ordinary physical interaction in which one system
becomes correlated with another. Still, not every interaction is already
a measurement. A measurement is a coupling that produces a sufficiently
stable, accessible, and record-like correlation. In realistic cases this
usually involves amplification, environmental stabilization, pointer-like
records, or decoherence-like mechanisms. This appeal to decoherence-like
stability does not by itself solve every version of the measurement
problem; it only clarifies why some correlations become usable as
records. The so-called measurement problem is therefore reinterpreted,
on the relational reading, as a problem about the compatibility and
translation of descriptions across \(P_{\mathrm{phys}}\)-perspectives,
not as the intrusion of consciousness into physics.

This point is central to the New Tractatus Program. If measurement is
ordinary coupling, then there is no need for a privileged observer
outside the world. Observers are systems. Instruments are systems.
Records are systems. What differs is the role they play in the network
of correlations.

Thus the classical picture

\[
    \text{system} + \text{external observer}
\]

is replaced by

\[
    \text{system} + \text{apparatus} + \text{interaction}
    + \text{stable record}.
\]

The observer is no longer a metaphysical exception. It is one physical
system among others, although a system may play the role of an observer
only when the relevant correlations become usable as records within a
given perspective.

% ---------------------------------------------------------
\subsection{The Quantum Revision of Wittgenstein}
% ---------------------------------------------------------

The result is a genuine revision of Wittgenstein, not merely a
Wittgensteinian decoration of quantum theory. The original
\emph{Tractatus} says:

\[
    \text{The world is the totality of facts.}
\]

\emph{Tractatus Quanticus} says, in effect:

\[
    \text{The world is the open network of facts from perspectives.}
\]

The difference is decisive. Wittgenstein denied that we could speak
from outside the world. Covoni and Rovelli argue, within the relational
interpretation, that quantum physics does not require a view from
outside all perspectives. The old limit of language becomes, on this
reading, a new limit of physical description.

This is why \emph{Tractatus Quanticus} can be read as the physical and
ontological starting point of the New Tractatus Program. It keeps
Wittgenstein's discipline against metaphysical excess, but replaces the
classical background with relational quantum mechanics. It does not
deny reality. It denies that, on the relational reading, reality must
appear as one complete, absolute, perspective-free totality.

The core argument may be summarized as follows:

\[
\begin{array}{rcl}
\text{world} &:& \text{not a totality of absolute facts;} \\[1mm]
\text{fact} &:& \text{a value of a variable in a perspective;} \\[1mm]
\text{perspective} &:& \text{a physical relation, not a mind;} \\[1mm]
\text{information} &:& \text{correlation between systems;} \\[1mm]
\text{measurement} &:& \text{ordinary coupling leaving records;} \\[1mm]
\text{silence} &:& \text{required where no perspective supplies information.}
\end{array}
\]

The next step is therefore philosophical. If the relational reading is
right that quantum theory does not support an absolute inventory of
facts, then ontology can be reconstructed in anti-totalitarian terms.

% =========================================================
\section{From Anti-Realism to Anti-Absolutism}
% =========================================================

A common misunderstanding of perspectival facts is to read them as
anti-realism. If facts are facts only from perspectives, one may think
that reality has been reduced to opinion, experience, or subjective
construction. This is not the position of the New Tractatus Program.
The point is not that reality disappears, but that reality is never
given as a completed, perspective-free inventory.

The central formula is therefore:

\[
    \text{Reality is not abolished; it is de-absolutized.}
\]

% ---------------------------------------------------------
\subsection{Why Perspectival Facts are not Mere Subjectivism}
% ---------------------------------------------------------

Perspectival facts are not private mental contents. In
\emph{Tractatus Quanticus}, a perspective is not essentially a mind,
a consciousness, or a human observer. It is a physical standpoint
constituted by interaction, correlation, and the values of variables
available relative to a system \cite{CovoniRovelli2026,Rovelli1996}.

Thus the claim

\[
    \text{facts are perspectival}
\]

does not mean

\[
    \text{facts are invented by subjects}.
\]

It means rather:

\[
    \text{facts obtain within physically defined relations}.
\]

A detector registering a value, an atom interacting with another atom,
or an observer reading an instrument are not examples of private
projection. They are examples of physical coupling. The fact is
relative, but the relation is real.

This distinction matters. Subjectivism makes reality depend on
belief. Perspectival realism makes facts depend on physical relations.
The first is psychological; the second is ontological.

% ---------------------------------------------------------
\subsection{The Difference between Relational Realism and Relativism}
% ---------------------------------------------------------

The New Tractatus Program is also not relativism in the weak sense that
anything may count as true. Relativism usually suggests that different
standpoints produce incompatible claims with no rational or physical
standard for comparison. Relational realism says something stronger and
more disciplined: facts are relative to perspectives, but perspectives
themselves are parts of the physical world.

This means that perspectives can be compared, coupled, stabilized, and
sometimes translated into one another. They are not arbitrary islands.
They are linked by interactions and by records. A fact relative to one
system may be represented, checked, or reconstructed from another
system, provided there is a physical channel connecting them.

The difference can be summarized as follows:

\[
\begin{array}{rcl}
\text{relativism} &:& \text{truth dissolves into standpoint;} \\[1mm]
\text{anti-realism} &:& \text{reality is reduced to representation;} \\[1mm]
\text{relational realism} &:& \text{facts are real, but relation-bound.}
\end{array}
\]

The program therefore belongs closer to perspectival and structural
forms of realism and perspectival accounts of causation than to anti-realism
\cite{Fine1986,Putnam1981,Price2007,LadymanRoss2007}.
It rejects the absolute standpoint, not the reality of the relations
described from within standpoints.

% ---------------------------------------------------------
\subsection{Reality without an Absolute Inventory}
% ---------------------------------------------------------

The deepest target, however, is the idea of an absolute inventory: a final list
of all facts as they are independently of every possible perspective.
This is the metaphysical image inherited from classical physics. The
world is imagined as a complete catalogue of determinate properties,
and observers merely discover parts of that catalogue.

On the relational reading, quantum mechanics blocks this image. Some
variables have values only relative to particular interactions. Some
questions cannot be answered without specifying the physical perspective
\(P_{\mathrm{phys}}\) and measurement context in which the answer is to
hold. The failure is not merely epistemic in the ordinary sense of
ignorance about a pre-existing value. Rather, within this interpretation,
the demand for a perspective-free value may itself be ill-formed.

The New Tractatus Program therefore replaces the image

\[
    \text{world} = \text{absolute inventory of facts}
\]

with

\[
    \text{world} = \text{network of physically situated facts}.
\]

This is anti-absolutism rather than anti-realism. It denies that reality must
take the form of a total catalogue. It does not deny that there are
facts, events, records, interactions, and stable structures.

% ---------------------------------------------------------
\subsection{Objectivity after the Loss of the God's-Eye View}
% ---------------------------------------------------------

If there is no God's-eye view, objectivity must be reconstructed rather
than presupposed. Objectivity no longer means access to reality from no
perspective. It means stability across perspectives, reliability of
records, repeatability of couplings, and coherence among physically
connected descriptions.

This suggests a post-classical notion of objectivity:

\[
    \text{objectivity} =
    \text{invariance and communicability across perspectives}.
\]

Stability here means not mere agreement of opinions, but repeatable
correlations, durable records, and invariant relations under changes of
observer or descriptive frame.

Such objectivity is weaker than the classical dream of a view from
nowhere, but stronger than subjectivism. It is weaker because it does
not claim to describe reality from outside all perspectives. It is
stronger because it remains constrained by interaction, probability,
measurement, and record.

The loss of the God's-eye view therefore does not leave us with chaos.
It leaves us with a more modest and more physical conception of truth.
Facts are not arbitrary; they are situated. Knowledge is not absolute;
it is embodied. Reality is not cancelled; it is de-absolutized.

The argument of this section may be condensed into four claims:

\[
\begin{array}{rcl}
\text{perspectivality} &\neq& \text{subjectivism;} \\[1mm]
\text{relational realism} &\neq& \text{relativism;} \\[1mm]
\text{no absolute inventory} &\neq& \text{no reality;} \\[1mm]
\text{no God's-eye view} &\neq& \text{no objectivity.}
\end{array}
\]

This prepares the next step. If reality is not a single closed
inventory, then the ontology suggested by the New Tractatus Program is
anti-totalitarian in a precise sense: it rejects the demand that the
world must be one final, complete, perspective-independent totality.
Here the relevant notion of perspective is primarily \(P_{\mathrm{phys}}\),
while objectivity requires stable translation into \(P_{\mathrm{epist}}\).

% =========================================================
\section{\emph{Against Totalitarianism}: The Attack on the Single World-Totality}
% =========================================================

Ismael and Price's \emph{Against Totalitarianism} gives the New
Tractatus Program its explicit philosophical name. Their text is not a
separate ontology but an interpretive introduction to
\emph{Tractatus Quanticus}. Its central function is to explain why the
quantum revision of Wittgenstein is also an attack on the idea of a
single, closed, perspective-independent world.

% ---------------------------------------------------------
\subsection{Why \enquote{Totalitarianism} is an Ontological Term Here}
% ---------------------------------------------------------

The word \enquote{totalitarianism} is not primarily political in this
context. It names an ontological temptation: the demand that reality
must form one final, complete totality of facts. On this picture, every
true description would ultimately belong to one master inventory.

Ismael and Price reject this demand. Their point is not that reality is
chaotic or arbitrary, but that the fantasy of one total view is a
metaphysical idol. On their relational reading, quantum mechanics does not merely limit
our access to a totality; it undermines the assumption that such a
totality is the right form for reality.

% ---------------------------------------------------------
\subsection{The World as a Closed Totality of Facts}
% ---------------------------------------------------------

The original \emph{Tractatus} already creates tension around totality.
It begins by identifying the world with the totality of facts, yet it
also denies us a position outside the world from which that totality
could be surveyed. Ismael and Price stress this point: the world as
whole is not one more object inside a larger space.

The classical temptation, however, survived in physics. Even if no
human observer could see the world from nowhere, classical physics
still suggested that there was a complete public landscape of objects
and intrinsic properties. The New Tractatus Program targets precisely
this residue of absolutism.

% ---------------------------------------------------------
\subsection{Quantum Mechanics against the Non-Relational Core}
% ---------------------------------------------------------

Ismael and Price's strongest claim is that the relational reading
interprets quantum mechanics as removing the non-relational core
imagined by classical physics. There is no additional classical inventory behind perspectives
stocked with intrinsic properties waiting to be discovered. Facts such
as position, momentum, or spin should not be attributed without
specifying the relevant interaction, system, and measurement context.
Their attribution depends on what has interacted with what.

The classical image was:

\[
    \text{many perspectives on one underlying world.}
\]

The quantum-perspectival image is:

\[
    \text{perspectives related to perspectives.}
\]

This is why their phrase \enquote{perspectives all the way down}
matters. It does not mean fantasy or subjectivism. It means that
physical reality is constituted through relations, correlations, and
interactions, not through an intrinsic inventory lying behind them.

% ---------------------------------------------------------
\subsection{Later Wittgenstein and the Plurality of Language-Games}
% ---------------------------------------------------------

Ismael and Price also connect the quantum lesson to later
Wittgenstein. The early \emph{Tractatus} still tends to treat language
as if its essential business were the representation of facts. The
later Wittgenstein rejects this unity. Language has many uses, many
games, and no single essence \cite{Wittgenstein1953}.

This pluralism reinforces the anti-totalitarian point. If language
itself has no single master-function, then the world should not be
forced into one master-description. The plurality of language-games and
the plurality of physical perspectives converge against the same
metaphysical fantasy: the fantasy of a single final form.

% ---------------------------------------------------------
\subsection{The Collapse of the Public/Private Metaphysical Divide}
% ---------------------------------------------------------

Finally, Ismael and Price argue that the perspectival view collapses
the old divide between public physics and private experience. Classical
physics placed intrinsic properties in the public world and appearances
in the private subject. Quantum perspectivalism removes the assumption
that made this division seem necessary: a perspective-independent
reality standing on one side, and subjective appearance on the other.

A perspective is not a private mental theatre. It is a physical system
in correlation with other systems: a detector, a thermometer, a brain,
or a particle. What one system can register about another is the
pattern of correlations embodied between them. Thus perspectives are
not sealed private worlds; they are physical relations that may become
accessible from further perspectives through suitable coupling and
records.

The argument may be summarized as follows:

\[
\begin{array}{rcl}
\text{totalitarian ontology} &:&
\text{one final world-totality;} \\[1mm]
\text{classical residue} &:&
\text{intrinsic properties behind appearances;} \\[1mm]
\text{quantum correction} &:&
\text{facts relative to physical perspectives;} \\[1mm]
\text{later Wittgenstein} &:&
\text{no single master-language;} \\[1mm]
\text{result} &:&
\text{no final classical inventory behind all perspectives.}
\end{array}
\]

This section therefore prepares the transition to consciousness. If
there is no final classical inventory behind all perspectives, then
consciousness should not be treated as a private exception to physics.
It should be understood as a special organization of perspective within
the physical world.

% =========================================================
\section{Anti-Totalitarian Ontology: Definition and Consequences}
% =========================================================

The central concept of the article can now be stated more directly. Anti-totalitarian
ontology is the rejection of the idea that reality must form one
closed, final, perspective-independent totality of facts. It is not a
denial of facts. It is a denial that all facts must belong to one
absolute catalogue.

Its guiding formula is:

\[
    \text{There are facts, but there is no final fact-of-all-facts.}
\]

This formula must be read in four distinct senses. Otherwise
anti-totalitarian ontology would remain an evocative slogan rather than
a disciplined thesis:
\[
\begin{array}{rcl}
\text{epistemic anti-totality} &:&
\text{no finite knower has access to all facts;} \\[1mm]
\text{semantic anti-totality} &:&
\text{no single master-language exhausts all descriptions;} \\[1mm]
\text{quantum anti-totality} &:&
\text{no global context-free assignment of all quantum values is assumed;} \\[1mm]
\text{ontological anti-totality} &:&
\text{no final perspective-independent inventory is presupposed.}
\end{array}
\]
The strongest claim of the article is the last one, but it is motivated
through the weaker and more technical claims preceding it. In particular,
the quantum claim is made from within the relational interpretation and
does not pretend to refute every realist interpretation of quantum theory.
In the broader foundations of quantum mechanics, however, this point is
also naturally related to contextuality results such as the
Kochen--Specker theorem, which rules out non-contextual global
value assignments for quantum systems of Hilbert-space dimension at
least three, under the standard requirement that functional relations
among compatible observables be preserved \cite{KochenSpecker1967}. The
present article uses that background only as support for the anti-totality
intuition; its positive interpretation remains relational.

% ---------------------------------------------------------
\subsection{What Anti-Totalitarian Ontology Means}
% ---------------------------------------------------------

Anti-totalitarian ontology means that reality is not exhausted by one
complete description from nowhere. Facts obtain, but they obtain within
perspectives, interactions, records, and contexts of attribution. The
world is therefore not a sealed totality placed before an impossible
observer. It is an open structure of physically situated facts.

In short, the view keeps realism while rejecting absolutism:

\[
    \text{realism without totality.}
\]

% ---------------------------------------------------------
\subsection{What It Does Not Mean: Not Politics, Not Chaos, Not Relativism}
% ---------------------------------------------------------

The term \enquote{totalitarianism} is used here ontologically, not as a
direct political category. It names the metaphysical demand for one
total description. The term is deliberately provocative, but its
function is diagnostic rather than political: it names the demand for a
single total form of reality, not any historical political regime. Nor
does anti-totalitarian ontology mean chaos. It does not say that
anything may count as real or true.

It is also not simple relativism. Relativism dissolves truth into
standpoint. Anti-totalitarian ontology instead says that standpoints
are themselves physically constrained. Perspectives are not arbitrary;
they are embodied in interactions, correlations, and records.

% ---------------------------------------------------------
\subsection{No Final Catalogue of Facts}
% ---------------------------------------------------------

The classical image of reality is a catalogue: every property, every
event, every fact has its place in one complete inventory. Quantum
perspectivalism breaks this image. Some facts are facts only relative
to particular interactions. Some variables have no value outside a
specified perspective.

The replacement is:

\[
    \text{absolute catalogue}
    \quad \longrightarrow \quad
    \text{network of situated facts}.
\]

Reality remains structured, but the structure is not a single completed
list.

% ---------------------------------------------------------
\subsection{No Privileged Metaphysical Observer}
% ---------------------------------------------------------

If there is no final catalogue, there is also no privileged observer
who could possess it. The metaphysical observer is the old fantasy of a
subject outside the world, surveying the world as a whole. Both
Wittgenstein's original critique and the quantum revision reject this
position.

Every observer is a system. Every description is made from within some
physical and conceptual situation. Objectivity therefore cannot mean
escape from perspective. It must mean stability, communicability, and
invariance across perspectives.

% ---------------------------------------------------------
\subsection{No Single Master-Language for Reality}
% ---------------------------------------------------------

Anti-totalitarian ontology also rejects the idea of one master-language
into which all meaningful descriptions must be translated. Physics,
phenomenology, biology, computation, and ordinary language do not all
serve one identical descriptive function.

This point echoes later Wittgenstein: language has many uses, and no
single essence \cite{Wittgenstein1953}. The plurality of
language-games supports the plurality of perspectives. Reality does not
require one final vocabulary in order to be real.

% ---------------------------------------------------------
\subsection{Ontology as Open, Layered, and Perspective-Bound}
% ---------------------------------------------------------

The consequence is an ontology that remains open, layered, and
perspective-bound. It is open because no final totality closes the
world. It is layered because different descriptions reveal different
structures. It is perspective-bound because facts are always articulated
from within some physical or conceptual standpoint.

This can be summarized as follows:

\[
\begin{array}{rcl}
\text{open} &:& \text{no final closure of reality;} \\[1mm]
\text{layered} &:& \text{many legitimate descriptive levels;} \\[1mm]
\text{perspective-bound} &:& \text{facts require conditions of attribution;} \\[1mm]
\text{realist} &:& \text{relations and records are physically constrained.}
\end{array}
\]

Anti-totalitarian ontology therefore gives the New Tractatus Program
its conceptual center. It preserves facts, but rejects the fantasy of a
single fact-of-all-facts. It preserves objectivity, but rejects the
view from nowhere. It preserves reality, but understands reality as
physically situated, relational, and open.

% =========================================================
\section{\emph{Tractatus de Conscientia}: Consciousness as Operational Perspective}
% =========================================================

\emph{Tractatus de Conscientia} extends the New Tractatus Program from
quantum ontology to consciousness. Its central move is to treat
consciousness neither as a mysterious extra substance nor as mere
outward behavior, but as a special organization of physical
perspective.

This proposal is continuous with, but not identical to, several major
families of contemporary consciousness theory. It shares with access
theories and Global Workspace approaches the emphasis on availability
for report, control, and flexible use \cite{Baars1988,Block1995,Dehaene2014};
with Integrated Information Theory it shares the idea that unity cannot
be reduced to a mere list of independent local states \cite{Tononi2004};
and with predictive-processing or free-energy approaches it shares the
importance of temporally organized control and model-updating
\cite{Friston2010}. At the same time, it keeps the hard-problem
pressure visible: phenomenal appearance cannot simply be identified
with reportability, information integration, or functional access
\cite{Chalmers1995}. These parallels should not be read as theoretical
equivalences. Global Workspace Theory, Integrated Information Theory,
and predictive-processing accounts make different explanatory commitments
and often disagree about what consciousness primarily consists in. The
aim is therefore not to replace those theories, but to state the
operational grammar within which claims about conscious perspective can
be made.

The crucial distinction is:

\[
    \text{all conscious perspectives are physical perspectives,}
\]

but

\[
    \text{not all physical perspectives are conscious perspectives.}
\]

% ---------------------------------------------------------
\subsection{From Physical Perspective to Conscious Perspective}
% ---------------------------------------------------------

\emph{Tractatus Quanticus} generalizes perspective: any physical system
in correlation with another may instantiate a perspective. \emph{Tractatus
de Conscientia} narrows the question. It asks when a physical
perspective becomes a conscious one.

A conscious perspective \(P_{\mathrm{conc}}\) requires more than
physical correlation \(P_{\mathrm{phys}}\). It requires structured
distinctions, integration, temporal stability, access to control, and
possible report or other evidential channels. A detector may register a
value, but it does not thereby possess a point of view.

These conditions should not yet be treated as jointly sufficient for
consciousness. Many non-conscious systems---thermostats with memory,
immune systems, recurrent artificial networks, distributed sensor arrays,
markets, or autonomous robots---may display integration, temporal
extension, and control-guiding behavior. The present proposal is
therefore a framework for necessary or diagnostic constraints, not a
completed sufficiency criterion. A full theory must still explain why
some integrated control systems instantiate \(P_{\mathrm{conc}}\), while
others remain merely physical, biological, or computational perspectives.

% ---------------------------------------------------------
\subsection{Appearance, Access, and Formal Structure}
% ---------------------------------------------------------

The argument separates three layers:

\[
\begin{array}{rcl}
\text{appearance} &:& \text{what occurs for an agent;} \\[1mm]
\text{access} &:& \text{what can guide report, control, or memory;} \\[1mm]
\text{formal structure} &:& \text{what remains invariant under redescription.}
\end{array}
\]

This separation prevents two errors. The first is mysticism: treating
appearance as a non-physical essence. The second is behaviorism:
identifying consciousness with external behavior alone. Consciousness
requires an operational bridge between internal organization and
accessible records.

The word \enquote{appearance} does not solve the hard problem by naming
a hidden object. It marks the first-person phenomenal aspect that
motivates the theory: what it is like for a system, if there is
something it is like. The claim of this article is more modest:
appearance is not reducible to access, but it can enter scientific
discussion only through access conditions, formal structure, and
evidential protocols. The article therefore does not explain phenomenal
appearance in reductive terms. It only asks under what conditions claims
about appearance can become scientifically tractable.

% ---------------------------------------------------------
\subsection{Conscious Episodes as Temporally Thick Regimes}
% ---------------------------------------------------------

A conscious episode is not an instant. It is a temporally extended
regime in which distinctions are maintained long enough to form a
stable present. The \enquote{now} of consciousness is therefore not a
mathematical point, but a window of integration.

This gives the theory a temporal condition stronger than the trivial
claim that processes take time:

\[
    \Delta \in [\Delta_{\min},\Delta_{\max}].
\]

The relevant interval must be fixed by the system's mode of integration:
memory, action, report, control, neural dynamics, or computational
update. It is not universal; it is system-relative and must be determined
by the architecture under study and by the evidential protocol used to
study it. A system without temporal retention may process signals, but
it lacks the stabilized continuity required for a conscious episode. A
system with only very slow integration, by contrast, may fail to form
the kind of unified present required for ordinary conscious access.

% ---------------------------------------------------------
\subsection{Integration: The Whole over the Parts}
% ---------------------------------------------------------

The unity of consciousness is treated as a whole-over-parts surplus. A
conscious episode is not a mere list of local distinctions. Its parts
must constrain one another so that the system forms one perspective.

A simplified integration surplus may be written as:

\[
    C_{\Delta}(t)
    =
    I(S_t;S_{t+\Delta})
    -
    \sum_{i=1}^{n}
    I\!\left(S^{(i)}_t;S^{(i)}_{t+\Delta}\right).
\]

If \(C_{\Delta}(t)>0\), the whole carries predictive or retentive
structure not reducible to independent part-wise transitions. This
quantity should be treated only as a schematic candidate diagnostic of
temporal integration, not as a validated measure of consciousness. Its
value depends on the chosen variables, the partition into parts, the
timescale \(\Delta\), and the background statistical model. It may be
positive because of common inputs, redundancy, or passive dependence
rather than consciousness itself. A more developed version would also
need to distinguish synergy, redundancy, and common-cause dependence,
for example by using tools in the spirit of partial information
decomposition \cite{WilliamsBeer2010}. It is also not invariant under
arbitrary coarse-graining or re-partitioning of the system, which limits
its role as a general measure. At most, \(C_{\Delta}(t)\) gives a formal
constraint that a fuller theory of conscious unity would have to refine.

% ---------------------------------------------------------
\subsection{Reportability, Control, Memory, and Records}
% ---------------------------------------------------------

Consciousness is not identical with report, but report matters because
science needs records. An internal distinction becomes operationally
relevant when it can affect speech, choice, action, memory, or another
publicly accessible channel.

Thus the theory avoids both extremes:

\[
    \text{unreported} \neq \text{absent},
\]

but also

\[
    \text{inaccessible} \neq \text{scientifically identifiable}.
\]

The point is not that consciousness must always be verbally reportable.
The point is that attribution requires some possible access route:
behavioral, neural, computational, physiological, or causal. This
matters for dreams, infants, non-human animals, anesthesia awareness,
locked-in states, and other cases in which speech is absent or
impaired. In such cases, report is only one possible evidential channel;
it is not the identity of consciousness itself.

% ---------------------------------------------------------
\subsection{The Self as Self-Index, Not Hidden Substance}
% ---------------------------------------------------------

The self is not treated as a metaphysical spectator behind experience.
It is a dynamical role within the system: a self-index that binds
episodes across time through memory, control, and prediction.

Formally, one may write:

\[
    Z_t = h(S_t),
\]

where \(Z_t\) is the internal variable or code that stabilizes the
system's updates across contexts. This notation is only a placeholder
for a future model of self-indexing; by itself it does not distinguish
a self from any ordinary internal state variable. The self is therefore
not a hidden object. It is an invariant of organization only if such a
role is specified by further structural and operational constraints.

This continues Wittgenstein's lesson: the self should not be imagined
as an object inside the world. In an operational theory, it appears as
a role performed by the system.

% ---------------------------------------------------------
\subsection{Measurement Limits and the End of Identifiability}
% ---------------------------------------------------------

The final consequence concerns measurement and attribution. Every attribution of
consciousness requires coupling to the system. But coupling is never
neutral: it selects channels, imposes conditions, and changes what can
be observed.

Therefore there is no protocol-free identifier of
\enquote{what-it-is-like}. Consciousness is not directly read off from
the world. It is inferred under explicit conventions, measurements, and
uncertainty bounds.

The resulting position is:

\[
\begin{array}{rcl}
\text{consciousness} &\neq& \text{mysterious substance;} \\[1mm]
\text{consciousness} &\neq& \text{mere behavior;} \\[1mm]
\text{consciousness} &=& \text{integrated, temporally thick, operational perspective.}
\end{array}
\]

This last line is a compressed programmatic formula, not a sufficiency
theorem. It identifies the target structure of \(P_{\mathrm{conc}}\),
while leaving open the additional constraints needed to distinguish
conscious integration from non-conscious control, regulation, or
information processing.

This makes \emph{Tractatus de Conscientia} the consciousness-theoretic
extension of the New Tractatus Program. It accepts the physicalization
of perspective, but adds the conditions needed for a perspective to
count as conscious: integration, access, memory, control, temporal
thickness, and measurable traces.

% =========================================================
\section{The Bridge Concept: Perspective}
% =========================================================

The concept of perspective is the bridge between Wittgenstein,
relational quantum mechanics, and consciousness theory. Yet it must be
used carefully. If \enquote{perspective} means everything, it explains
nothing. The central claim is therefore:

\begin{quote}
\emph{Tractatus Quanticus} generalizes perspective; \emph{Tractatus de
Conscientia} restricts and enriches it.
\end{quote}

% ---------------------------------------------------------
\subsection{Perspective in Wittgenstein}
% ---------------------------------------------------------

In Wittgenstein, perspective is not yet a technical physical concept.
It appears indirectly through the limits of language, world, and self.
The subject does not stand outside the world and look at it as an
object. Rather, the subject is the limit from which a world is given
\cite{Wittgenstein1922}.

Thus the original problem is already perspectival:

\[
    \text{there is no description from nowhere.}
\]

But Wittgenstein does not yet physicalize perspective. That move belongs
to the quantum revision.

% ---------------------------------------------------------
\subsection{Perspective in Relational Quantum Mechanics}
% ---------------------------------------------------------

In relational quantum mechanics, perspective means primarily
\(P_{\mathrm{phys}}\): a physical relation. It is not a private
consciousness, but a system's position within a network of interactions,
correlations, and records \cite{Rovelli1996,CovoniRovelli2026}.

A fact such as

\[
    A = a
\]

is therefore not absolute. It is a value of a variable relative to a
system or interaction. Perspective is generalized beyond human
observers:

\[
    P_{\mathrm{phys}} = \text{physical correlation structure}.
\]

This is the decisive expansion made by \emph{Tractatus Quanticus}.

% ---------------------------------------------------------
\subsection{Perspective in Consciousness Theory}
% ---------------------------------------------------------

\emph{Tractatus de Conscientia} accepts the physicalization of
perspective but adds further requirements. A conscious perspective \(P_{\mathrm{conc}}\) is
not merely any physical relation \(P_{\mathrm{phys}}\). It requires
appearance, access, integration, temporal thickness, memory, control,
and possible records \cite{SienickiSienicki2026}.

Thus consciousness is not added to physics from outside. It is a
special organization within physical perspective:

\[
    \text{conscious perspective}
    =
    \text{integrated and operationally accessible physical perspective}.
\]

% ---------------------------------------------------------
\subsection{Physical Perspective versus Conscious Perspective}
% ---------------------------------------------------------

The distinction can be stated simply:

\[
    \text{all conscious perspectives are physical perspectives,}
\]

but

\[
    \text{not all physical perspectives are conscious perspectives.}
\]

A detector may instantiate a physical perspective because it becomes
correlated with another system. But a detector does not thereby possess
a conscious point of view. Consciousness requires richer organization:
structured distinctions, unified integration, temporal continuity, and
access to control or report.

The distinction is therefore:

\[
\begin{array}{rcl}
\text{physical perspective} &:&
\text{correlation, interaction, variable-value attribution;} \\[1mm]
\text{conscious perspective} &:&
\text{integration, access, temporal thickness, self-indexing.}
\end{array}
\]

% ---------------------------------------------------------
\subsection{Why This Distinction Matters}
% ---------------------------------------------------------

This distinction prevents two errors. The first error is panpsychic
inflation: if every physical perspective were conscious, then
consciousness would become too cheap. The second error is dualist
exceptionalism: if consciousness were not physical perspective at all,
then it would become mysterious again.

The New Tractatus Program avoids both:

\[
\begin{array}{rcl}
\text{\emph{Tractatus Quanticus}} &:&
\text{perspective is generalized;} \\[1mm]
\text{\emph{Tractatus de Conscientia}} &:&
\text{perspective is enriched;} \\[1mm]
\text{result} &:&
\text{consciousness is physical, but not every physical relation is conscious.}
\end{array}
\]

Perspective is therefore the bridge concept, but only if it remains
differentiated. It connects world, fact, measurement, and consciousness
without collapsing them into one vague term.

% =========================================================
\section{Critical Assessment: Promise and Risk of the New Tractatus Program}
% =========================================================

The New Tractatus Program is powerful because it gives a common
grammar for problems usually kept apart: quantum facts, ontology,
measurement, selfhood, and consciousness. Yet the same ambition also
creates risks. The program must avoid becoming a general metaphor in
which every problem is solved by saying \enquote{perspective}. Its
strength depends on keeping distinctions sharp.

% ---------------------------------------------------------
\subsection{The Promise: A Unified Grammar for Physics and Consciousness}
% ---------------------------------------------------------

The main promise of the program is unification without reduction. It
does not reduce consciousness to quantum mechanics, nor quantum
mechanics to consciousness. Instead, it identifies a shared grammar:
facts, perspectives, couplings, records, and limits of attribution.

This allows one to compare three domains:

\[
\begin{array}{rcl}
\text{quantum mechanics} &:& \text{facts relative to physical systems;} \\[1mm]
\text{ontology} &:& \text{reality without a final totality;} \\[1mm]
\text{consciousness} &:& \text{integrated, operationally accessible perspective.}
\end{array}
\]

The gain is conceptual economy. The same anti-absolutist discipline
applies to world, fact, measurement, and self.

% ---------------------------------------------------------
\subsection{The Risk: Inflation of the Word \enquote{Perspective}}
% ---------------------------------------------------------

The main risk is inflation. If every relation, every description, every
system, and every conscious episode is called a perspective, the term
loses explanatory value. It becomes a slogan rather than a concept.

The solution is differentiation. At minimum, the article must separate:

\[
\begin{array}{rcl}
\text{physical perspective} &:& \text{interaction and correlation;} \\[1mm]
\text{epistemic perspective} &:& \text{information available to a knower;} \\[1mm]
\text{biological perspective} &:& \text{organism-relative perception and action;} \\[1mm]
\text{conscious perspective} &:& \text{integrated, accessible, temporally thick experience.}
\end{array}
\]

Without this taxonomy, the New Tractatus Program risks collapsing into
conceptual vagueness. Later uses of the word \enquote{perspective}
should therefore be read with an implicit subscript: \(P_{\mathrm{phys}}\)
for physical correlations, \(P_{\mathrm{epist}}\) for accessible
information, \(P_{\mathrm{bio}}\) for organism-relative action, and
\(P_{\mathrm{conc}}\) for conscious experience.

% ---------------------------------------------------------
\subsection{Does the Program Dissolve Too Much?}
% ---------------------------------------------------------

A second risk is excessive dissolution. The program often treats old
problems as pseudo-problems: the view from nowhere, the absolute
inventory of facts, the hidden self, the private essence of
consciousness. This is therapeutically useful, but it can become too
quick.

Some problems may be badly formulated, yet still point to real
phenomena. The hard problem of consciousness, for example, may contain
confused metaphysics, but it also marks the real gap between access,
report, and appearance. The task is therefore not simply to dissolve
the problem, but to reconstruct it in operational terms.

The rule should be:

\[
    \text{dissolve false grammar, but preserve real phenomena.}
\]

% ---------------------------------------------------------
\subsection{Can Operational Consciousness Avoid Behaviorism?}
% ---------------------------------------------------------

\emph{Tractatus de Conscientia} rejects behaviorism, because it does
not identify consciousness with outward behavior. Yet any operational
theory faces the danger of drifting back toward behaviorism if it
treats only reports and records as real.

The solution is to distinguish access from appearance. Reports are
evidence, not identity. Behavior is a channel, not consciousness
itself. A conscious episode may exceed what is reported, but scientific
attribution still requires some possible bridge to evidence.

Thus the position is:

\[
    \text{appearance is not reducible to report,}
\]

but

\[
    \text{science needs report, control, memory, or other records.}
\]

The position therefore has to walk a narrow path between mysticism and behaviorism.

% ---------------------------------------------------------
\subsection{Can Anti-Totalitarian Ontology Preserve Objectivity?}
% ---------------------------------------------------------

Another objection concerns objectivity. If there is no final catalogue
of facts, does objectivity disappear? The answer must be no. The
program preserves objectivity by relocating it.

Objectivity no longer means access to reality from nowhere. It means
stability across perspectives, reliable coupling, repeatable records,
and invariance under transformations of description.

Thus:

\[
    \text{objectivity}
    =
    \text{stable inter-perspectival constraint}.
\]

Here, again, stability means repeatable coupling, durable records,
communicability, and invariance under relevant changes of observer or
descriptive frame, not mere convergence of subjective opinion.

This is weaker than classical absolutism, but stronger than relativism.
It gives us objectivity without metaphysical totality.

% ---------------------------------------------------------
\subsection{The Need for Formal Development}
% ---------------------------------------------------------

Finally, the program still requires further formal development. Its central
concepts---perspective, fact, access, integration, reportability,
self-indexing, and identifiability---must be made technically sharper.

Several tasks follow:

\[
\begin{array}{rcl}
\text{physics} &:& \text{formalize perspective-dependent fact attribution;} \\[1mm]
\text{information theory} &:& \text{clarify mutual information, records, synergy, and redundancy;} \\[1mm]
\text{ontology} &:& \text{separate epistemic, semantic, quantum, and ontological anti-totality;} \\[1mm]
\text{consciousness theory} &:& \text{define integration, access, timescale, and anti-overgeneration conditions;} \\[1mm]
\text{methodology} &:& \text{state limits of identifiability and protocol design.}
\end{array}
\]

Without such development, the New Tractatus Program remains a powerful
philosophical orientation rather than a mature research framework.

The critical balance can therefore be stated as follows:

\[
\begin{array}{rcl}
\text{promise} &:& \text{a shared grammar of perspective and coupling;} \\[1mm]
\text{risk} &:& \text{conceptual inflation and premature dissolution;} \\[1mm]
\text{requirement} &:& \text{clear distinctions and formal constraints.}
\end{array}
\]

The program is strongest when it resists the temptation to say that
everything is perspective in the same sense. Its real contribution is more precise:
different kinds of perspective structure different kinds of fact,
objectivity, and consciousness.

% =========================================================
\section{Coupling, Evidence, and Silence}
% =========================================================

The three texts share a discipline of limits. Facts, measurements, and
attributions of consciousness do not float free of perspective. They
require coupling, channels, records, and conditions of access. This is
where the old Wittgensteinian theme of silence returns in operational
form.

% ---------------------------------------------------------
\subsection{Every Measurement is an Interaction}
% ---------------------------------------------------------

In \emph{Tractatus Quanticus}, measurement is not an exception to
physics. It is an ordinary physical interaction in which one system
becomes correlated with another \cite{CovoniRovelli2026}. A fact is not
revealed from outside the world; it is established within a relation.

Thus measurement has the form:

\[
    \text{system} + \text{apparatus} + \text{interaction}
    + \text{stable record}.
\]

There is no detached observer who simply reads reality as it is in
itself. Every observation belongs to a physical interaction.

% ---------------------------------------------------------
\subsection{Every Attribution of Consciousness Requires a Protocol}
% ---------------------------------------------------------

The same principle applies to consciousness. \emph{Tractatus de
Conscientia} argues that consciousness cannot be identified by a
protocol-free inspection of private experience. Every attribution
requires some operational route: report, behavior, neural signature,
memory trace, causal intervention, or computational record
\cite{SienickiSienicki2026}.

This does not reduce consciousness to report. Rather, it says that
scientific attribution requires access:

\[
    \text{consciousness claim}
    \quad \Rightarrow \quad
    \text{specified evidential protocol}.
\]

Without a protocol, the claim may be meaningful existentially, but it is
not scientifically identifiable.

% ---------------------------------------------------------
\subsection{Privacy as a Property of Coupling}
% ---------------------------------------------------------

Privacy is therefore not a metaphysical wall. It is a feature of
coupling. An experience is private when its internal distinctions are
not directly accessible to another system except through limited
channels.

The private/public distinction becomes physical:

\[
\begin{array}{rcl}
\text{private} &:& \text{weak, indirect, or unavailable coupling;} \\[1mm]
\text{public} &:& \text{stable, repeatable, shareable coupling and records.}
\end{array}
\]

This preserves the reality of first-person appearance without turning
it into a non-physical substance.

% ---------------------------------------------------------
\subsection{What No Specified Protocol Can Distinguish}
% ---------------------------------------------------------

A theory must also say where identifiability ends. It is too strong to
claim that science can know, in advance, that no possible future
protocol could distinguish two alleged states. The more defensible
principle is methodological: if no specified physically possible
protocol can distinguish two attributions, then there is no current
scientific basis for treating them as operationally distinct. This is
not a denial that something is experienced. It is a limit on what can
be claimed, tested, or compared under explicit evidential conditions.

The rule is:

\[
    \text{no specified physically possible protocol}
    \quad \Rightarrow \quad
    \text{no current scientific distinction in attribution}.
\]

This is the consciousness-theoretic version of the Tractarian boundary
between sense and pseudo-question.

% ---------------------------------------------------------
\subsection{Silence as Epistemic Discipline, Not Mysticism}
% ---------------------------------------------------------

Silence, in this program, is not mystical reverence for the ineffable.
It is epistemic discipline. One should speak where there are
constraints, measure where there are channels, infer where there is
evidence, and remain silent where no specified protocol can distinguish
the claim.

The discipline can be put in the following compressed form:

\[
\begin{array}{rcl}
\text{speak} &:& \text{where concepts have constraints;} \\[1mm]
\text{measure} &:& \text{where channels exist;} \\[1mm]
\text{infer} &:& \text{where evidence is available;} \\[1mm]
\text{remain silent} &:& \text{where no specified protocol can distinguish the claim.}
\end{array}
\]

This unites the three texts. Wittgenstein gives the grammar of limits;
\emph{Tractatus Quanticus} physicalizes those limits through
perspectival facts; \emph{Tractatus de Conscientia} operationalizes
them for consciousness. Silence is no longer simply the end of speech.
It is the point at which perspective, coupling, and evidence run out.

% =========================================================
\section{Conclusion: A Grammar of Partiality}
% =========================================================

The New Tractatus Program begins from Wittgenstein's old problem but
transforms it under post-classical conditions. The issue is no longer
only what language can say about the world. It is what can be said,
measured, or inferred from within a world of perspectives.

% ---------------------------------------------------------
\subsection{From the Totality of Facts to Facts from Perspectives}
% ---------------------------------------------------------

The original \emph{Tractatus} defined the world as the totality of
facts. The new program keeps the priority of facts, but rejects their
absolutization. A fact is not a free-standing item in a final catalogue.
It is what is the case from a perspective, under conditions of
interaction and attribution.

Thus the central movement is:

\[
    \text{totality of facts}
    \quad \longrightarrow \quad
    \text{facts from perspectives}.
\]

Reality is preserved, but it is no longer imagined as a completed
inventory visible from nowhere.

% ---------------------------------------------------------
\subsection{From the View from Nowhere to Embodied Physical Viewpoints}
% ---------------------------------------------------------

The second movement concerns the observer. The classical fantasy of a
view from nowhere is replaced by embodied physical viewpoints.
Observers are not outside the world. They are systems within it.
Measurements are not neutral openings onto absolute reality; they are
couplings between physical systems.

Objectivity therefore survives, but in a different form. It is no
longer the possession of an impossible total view. It is the stability
of relations, records, and invariances across perspectives.

\[
    \text{objectivity}
    =
    \text{stable inter-perspectival constraint}.
\]

% ---------------------------------------------------------
\subsection{From Consciousness as Mystery to Consciousness as Operationally Constrained Perspective}
% ---------------------------------------------------------

The third movement concerns consciousness. Consciousness is not treated
as a metaphysical exception to physics, nor as mere outward behavior.
It is understood as a special kind of physical perspective: integrated,
temporally thick, accessible to control and memory, and in some cases
reportable.

The crucial distinction remains:

\[
    \text{all conscious perspectives are physical perspectives,}
\]

but

\[
    \text{not all physical perspectives are conscious perspectives.}
\]

This prevents both dualism and inflation. Consciousness is physical,
but not every physical relation is conscious.

% ---------------------------------------------------------
\subsection{The New Tractatus Program as Post-Classical Philosophy}
% ---------------------------------------------------------

The New Tractatus Program is therefore post-classical in three senses.
Here \enquote{post-classical} means not anti-scientific or anti-realist,
but post-absolute: after the loss of the absolute state, the absolute
observer, and the absolute inventory of facts.
It is post-classical in physics, because it rejects the absolute
inventory of intrinsic properties. It is post-classical in ontology,
because it refuses the single closed world-totality. It is
post-classical in consciousness theory, because it replaces the hidden
self or ineffable essence with operationally constrained perspective.

Its final grammar may be summarized as follows:

\[
\begin{array}{rcl}
\text{world} &:& \text{not a closed totality, but an open field of perspectives;} \\[1mm]
\text{fact} &:& \text{not absolute, but perspective-bound;} \\[1mm]
\text{observer} &:& \text{not external, but physically situated;} \\[1mm]
\text{consciousness} &:& \text{not mysterious substance, but a constrained integrated perspective;} \\[1mm]
\text{silence} &:& \text{not mysticism, but discipline where evidence ends.}
\end{array}
\]

The program should not be presented as solving every problem. It requires further formal
development, especially in distinguishing physical, epistemic,
biological, and conscious perspectives. Yet its philosophical value is
already clear. It offers a grammar of partiality: a way of speaking
about reality without pretending to speak from outside it.

The New Tractatus Program does not deny reality; it denies that reality
must be available as a single, final, perspective-free totality.
% =========================================================
% REFERENCES
% =========================================================

\end{document}